\documentclass[preprint,prd,aps,article,nofootinbib]{revtex4}
\usepackage{graphicx}
\usepackage{diagbox}
\usepackage{float}
\usepackage[colorlinks,
            linkcolor=black,
            anchorcolor=blue,
            citecolor=black
            ]{hyperref}
\usepackage{doi}
\usepackage{color}
\usepackage{slashed}
\begin{document}
\title{$\eta_{_{c2}}(^1D_{_2})$ and its electromagnetic decays}
\author{Xin-Yao Du$^{1,2,3}$\footnote{duxinyao0401@163.com},
Su-Yan Pei$^{1,2,3}$,
Wei Li$^{1,2,3}$,
Man Jia$^{1,2,3}$,
Qiang Li$^{4}$,
Tianhong Wang$^{5}$,
Bo Wang$^{1,2,3}$,
Guo-Li Wang$^{1,2,3}$\footnote{wgl@hbu.edu.cn, corresponding author}}

\affiliation{${^1}$ Department of Physics, Hebei University, Baoding 071002, China
\nonumber\\$^{2}$ Hebei Key Laboratory of High-precision Computation and Application of Quantum Field Theory, Baoding 071002, China
\nonumber\\$^3$ Hebei Research Center of the Basic Discipline for Computational Physics, Baoding 071002, China
\nonumber\\$^4$School of Physical Science and Technology,  Northwestern Polytechnical University, Xi'an 710129, China
\nonumber\\$^5$ School of Physics, Harbin Institute of Technology, Harbin 150001, China
}
\begin{abstract}
The spin-singlet state $\eta_{_{c2}}(^1D_{_2})$ has not been discovered in experiment and it is the only missing low-excited $D$-wave charmonium, so in this paper, we like to study its properties. Using the Bethe-Salpeter equation method, we obtain its mass as $3828.2$ MeV and its electromagnetic decay widths as $\Gamma[\eta_{_{c2}}(1D)\rightarrow h_{_{c}}(1P)\gamma]=284$ keV, $\Gamma[\eta_{_{c2}}(1D)\rightarrow J/\psi\gamma]=1.04$ keV, $\Gamma[\eta_{_{c2}}(1D)\rightarrow\psi(2S)\gamma]=3.08$ eV, and $\Gamma[\eta_{_{c2}}(1D)\rightarrow\psi(3770)\gamma]=0.143$ keV. {Considering the strong decay widths are estimated to be $\Gamma(\eta_{_{c2}}(1D)\to\eta_c \pi\pi)=144~\rm{keV}$ and
$\Gamma(\eta_{_{c2}}(1D)\to gg)= 46.1~\rm{keV}$, we obtain the total decay width of $475$ keV for $\eta_{_{c2}}(1D)$, and point out that the full width is very sensitive to the mass $M_{\eta_{_{c2}}}$.} In our calculation, the emphasis is put on the relativistic corrections. Our results show that $\eta_{_{c2}}\rightarrow h_{_{c}}\gamma$ is the nonrelativistic $E1$ transition dominated $E1+M2+E3$ decay, and $\eta_{_{c2}}\rightarrow \psi\gamma$ is the $M1+E2+M3+E4$ decay but the relativistic $E2$ transition contributes the most.

\end{abstract}
\maketitle

\section{Introduction}

In the last two decades, the Belle, BABAR, BESIII, LHCb collaborations, etc.,  have discovered a large number of new charmoniumlike states \cite{Belle:2003nnu,Belle:2005rte,BABAR:2010jfn,BABAR:2012nxg,Belle:2017egg,Belle:2005lik,Belle:2007woe,BABAR:2005hhc,BESIII:2013ris,BESIII:2020qkh} that are above the charmonium threshold and many of which do not match with the predictions of the underlying model calculations based on the conventional  hadron assumptions \cite{godfrey1985, Barnes:2005pb}. These exotic hadrons are often referred to as $``X, Y, Z"$ particles, and have aroused great research interest. However, there are still some conventional low-excited charmonia that have not been discovered by experiments, such as the $\eta_{_{c2}}(1^1D_{_2})$ and $\chi_{_{c2}}(1^3F_{_2})$, etc.
The supplementation of these traditional mesons in the charmonium family is crucial for the success of traditional potential models and QCD theory.

In the charmonium family, there are four $D$-wave states, namely $\psi({}^3D_1)$, $\psi_2({}^3D_2)$, $\psi_3({}^3D_3)$, and $\eta_{_{c2}}({}^1D_{_2})$, their quantum numbers are $1^{--}$, $2^{--}$, $3^{--}$ and $2^{-+}$, respectively. $\psi({1}^3D_1)$, also known as the $\psi(3770)$, was the first observed $D$ wave charmonium because it can be produced directly in $e^{+}e^{-}$ collisions \cite{Rapidis:1977cv}. Belle \cite{Belle:2013ewt} and BESIII \cite{BESIII:2015iqd} observed the particle $X(3823)$ by its transition $X(3823)\to \gamma\chi_{_{c1}}$ in $B$ decay and $e^{+}e^{-}$ annihilation, respectively. And it is found that $X(3823)$ is a good candidate of $\psi_2({1}^3D_2)$ \cite{liwei1,cheny,Deng:2016stx}. There is also a good candidate for $\psi_3({1}^3D_3)$ \cite{liwei2,wangzg}, which is the recently observed particle $X(3842)$ in experiment \cite{LHCb2019}.
Currently, $\eta_{_{c2}}(1^1D_{_2})$ is the only missing low-excited $D$ wave charmonium. Therefore, in this article, we will study the properties of the  $\eta_{_{c2}}(1^1D_{_2})$.

In 2003, a narrow resonant state $X(3872)$ was discovered by the Belle collaboration  \cite{Belle:2003nnu}, and $\eta_{_{c2}}(1^1D_{_2})$ was once assigned to this particle. However, theoretical calculations strongly contradict experimental data on the electromagnetic decay of $X(3872)$ \cite{kalashnikova,Jia:2010jn,Ke:2011jf}, and the quantum number of $X(3872)$ was finally determined as $J^{PC}=1^{++}$ \cite{LHCb:2013kgk,LHCb:2015jfc}. In 2020, Belle Collaboration searched for $\eta_{_{c2}}(1^1D_{_2})$ within the mass range of 3795 $\sim$ 3845 MeV in channels of $B^{+}\rightarrow\eta_{_{c2}}(1^1D_{_2})K^{+}$, $B^{0}\rightarrow\eta_{_{c2}}(1^1D_{_2})K^{0}_{S}$, $B^{0}\rightarrow\eta_{_{c2}}(1^1D_{_2})\pi^{-}K^{+}$ and $B^{+}\rightarrow\eta_{_{c2}}(1^1D_{_2})\pi^{+}K^{0}_{S}$ \cite{Belle:2020esr}. After that, Belle Collaboration also looked for $\eta_{_{c2}}(1^1D_{_2})$ in the mass range of 3.8 $\sim$ 3.88 GeV in the process of $e^{+}e^{-}\rightarrow\gamma\eta_{_{c2}}(1^1D_{_2})$ \cite{Belle:2021cjk}. Unfortunately, no $\eta_{_{c2}}(1^1D_{_2})$ signal was observed.

In theory, $\eta_{_{c2}}(1^1D_{_2})$ is estimated to have a mass of 3.80 to 3.88 GeV \cite{Yang:2020pyh}, between the $D\bar{D}$ and $D^{\ast}\bar{D}$ thresholds. As can be seen from parity conservation, its decay into $D\bar{D}$ is forbidden. Therefore, $\eta_{_{c2}}(1^1D_{_2})$ is a narrow resonance state, and its main decay modes are strong decays to $gg$, $\eta_c\pi\pi$ \cite{Novikov:1977dq,Eichten:2002qv,Fan:2009cj,wth2013} and electromagnetic (EM) decays \cite{Ebert:2002pp,Barnes:2005pb}. It is found that the process of $\eta_{_{c2}}(1^1D_{_2})\rightarrow h_{_{c}}(1P)\gamma$ has the biggest branching ratio, so studying its EM radiative transitions is crucial for revealing the properties of $\eta_{_{c2}}(^1D_{_2})$. And the EM decays of $\eta_{_{c2}}(1^1D_{_2})$ have been studied by many models, such as the potential models \cite{Sebastian:1996cy,Eichten:2002qv,Barnes:2005pb}, the light front quark model \cite{Ke:2011jf}, the lattice QCD \cite{Yang:2012mya}, and the nonrelativistic models \cite{Li:2009zu,Jia:2010jn,Guo:2014zva,Deng:2016stx}, etc. However, in existing studies, people mainly focus on the greatest contribution, such as the electric dipole $E1$ transition, and relativistic correction has not been carefully considered. And we have found that the relativistic corrections are relatively large for $P$-wave and $D$-wave charmonia, which cannot be ignored \cite{wgl2020}.

So in this paper, we will provide a relativistic study on the radiative EM decays of $\eta_{_{c2}}(^1D_{_2})$ adopting the instantaneous Bethe-Salpeter (BS) method. The BS equation is a relativistic dynamic equation used to describe a two-body bound state in quantum field theory \cite{Salpeter:1951sz}, and its instantaneous approximation is the Salpeter equation \cite{Salpeter:1952ib} which is suitable for heavy mesons. In our method, we  construct the universal wave function of a meson according to its quantum number of $J^{P(C)}$, where 4 or 8 radial wave functions are unknown and are obtained numerically by solving the complete Salpeter equation. The relativistic wave function obtained in this way contains rich information, besides the main waves, all the meson wave functions contain other partial waves \cite{Wang:2022cxy}, which mainly contribute to the relativistic corrections. We will study the contributions of all the partial waves in detail in the main text. Besides the relativistic corrections, our method has been proven effective in many aspects, such as being able to explain the `$1/2$ vs $3/2$' puzzle \cite{Wang:2022tuy}. In the study of radiative EM transitions, this method is also very powerful \cite{Wang:2010ej,Wang:2015yea} because we not only calculate the leading order contribution, but also provide a complete calculation that includes the transitions of $E1+M2+E3+...$ or $M1+E2+M3+...$ \cite{liwei1,liwei2,peisy}.

Another special benefit of this method is that we can provide the correct mass splitting between the $2^{--}$ and $2^{-+}$ states. Since the mass of the $2^{--}$ state $\psi_2({1}^3D_2)$ has already been detected in the experiment, using it as an input parameter, we can provide a relatively reliable mass prediction of $2^{-+}$ state $\eta_{_{c2}}(1^1D_{_2})$. Therefore, in this article, we will study the mass spectra of $2^{-+}$ states as well as the $2^{--}$ states. For the radiative EM decays of $\eta_{_{c2}}(1^1D_{_2})$, we will focus on the main decay mode $\eta_{_{c2}}(1^1D_{_2})\rightarrow h_{_{c}}(1P)\gamma$, in addition, the decays to $\psi(1S)\gamma$, $\psi(2S)\gamma$ and $\psi(3770)\gamma$ final states are also calculated, and the full decay width of $\eta_{_{c2}}(1^1D_{_2})$ is estimated. The relativistic corrections and behaviors of different partial waves are discussed. Finally, we present the behavior of the decay widths of $\eta_{_{c2}}(1^1D_{_2})$ as a function of its mass in the range of 3800$\sim$3872 MeV.

This paper is organized as follows, In Sec. II, we first show the method of calculating the transition amplitude of the EM decay. Then we provide the relativistic wave functions used in this paper, including their diagrams. The mass spectra of $2^{--}$ and $2^{-+}$ charmonia as well as the form factors are also given in this section. In Sec. III, we give the results of the EM decays, the discussions and conclusion.

\section{THE THEORETICAL CALCULATIONS}

\subsection{Transition amplitude of EM decay}

\begin{figure}[!htb]
\begin{minipage}[c]{1\textwidth}
\includegraphics[width=3in]{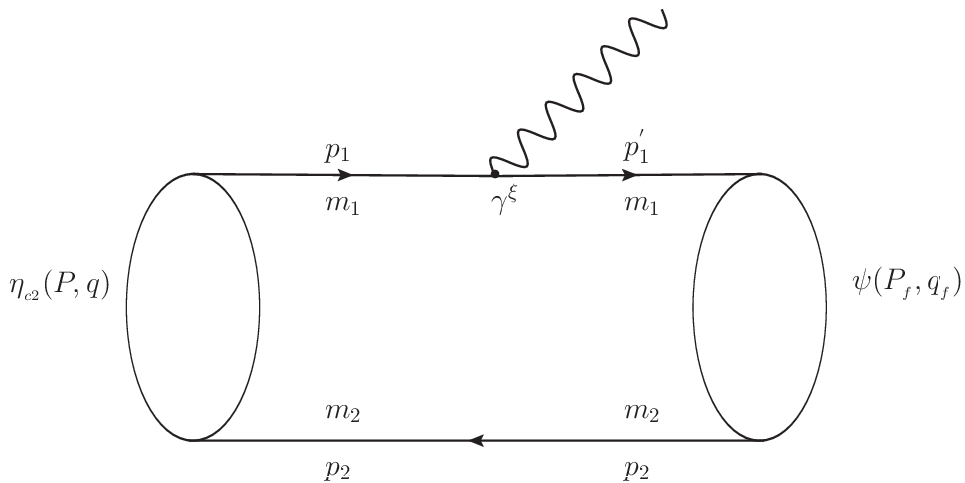}
\includegraphics[width=3in]{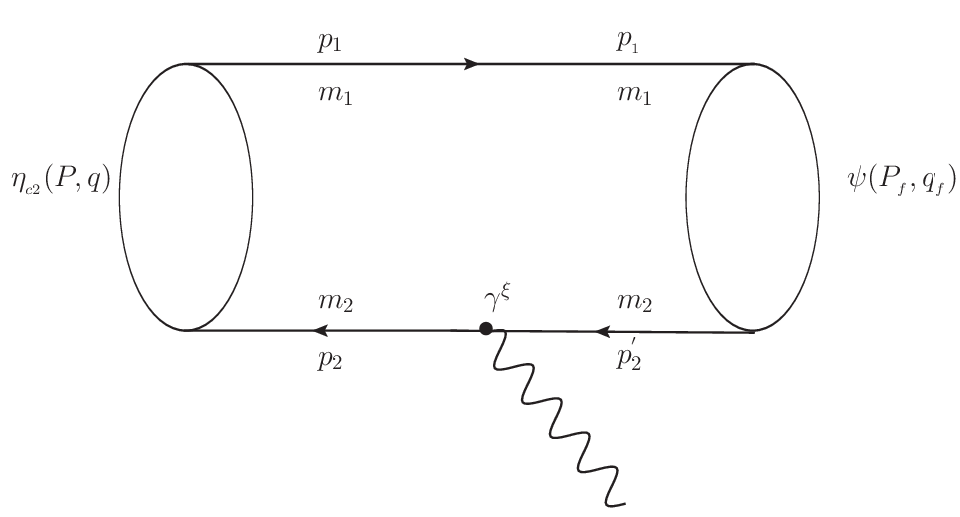}
\end{minipage}
\caption{Feynman diagrams for the EM transition of $\eta_{_{c2}}\rightarrow\psi\gamma$.}
\label{feynman}
\end{figure}

We take the decay channel $\eta_{_{c2}}(^1D_{_2})\rightarrow\psi(^3S_{_1})\gamma$ as an example to show our method how to calculate the transition amplitude. In Figure \ref{feynman}, we show the Feynman diagrams for the radiative EM transition of $\eta_{_{c2}}\rightarrow\psi\gamma$, and the corresponding transition amplitude can be written as
\begin{eqnarray}
&&\langle \psi(P_{_f},\epsilon_{_f})\gamma(k,\epsilon_{_0})|\eta_{_{c2}}(P,\epsilon_{_i})\rangle=(2\pi)^{4}\delta^{4}
(P-P_{_f}-k)\epsilon_{_{0{\xi}}}{\cal M}^{\xi},
\end{eqnarray}
where $\epsilon_{_0}$ is the polarization vector of the emitting photon, while $\epsilon_{_i}$ and $\epsilon_{_f}$ are the polarization vectors of the initial and final mesons, respectively. $P$, $P_{f}$ and $k$ are the momenta of the initial meson, final meson and final photon, respectively.

Fig. \ref{feynman} shows that the Feynman diagrams of EM transition consist of two parts, where photons are emitted from quark and antiquark, respectively. And the corresponding hadronic matrix element ${\cal M}^{\xi}$ can be written as the overlapping integral over the wave functions of initial and final mesons \cite{zhangzx}
\begin{eqnarray}\label{em}
&&{\cal M}^{\xi}=\int{\frac{d^3q_{_\perp}}{(2\pi)^{3}}}Tr\bigg[Q_{_1}e\frac{\slashed{P}}{M}\bar{\varphi}^{++}_{_f}
(q_{_\perp}+\alpha_{_2}P_{_{f_\perp}})
\gamma^{\xi}\varphi^{++}_i(q_{_\perp})
\nonumber\\&&\hspace{1.3cm}+Q_{_2}e~\bar{\varphi}^{++}_{_f}(q_{_\perp}-\alpha_{_1}P_{_{f_\perp}})
\frac{\slashed{P}}{M}\varphi^{++}_{_i}(q_{_\perp})
\gamma^{\xi}\bigg],
\end{eqnarray}
where $q$ is the relative momentum between quark and antiquark. We have defined $q_{_\perp}=q-\frac{P\cdot q}{M^{2}}P$ and $P_{_{f_\perp}}=P_{_{f}}-\frac{P\cdot P_{_{f}}}{M^{2}}P$. $\alpha_{_i}=\frac{m_{i}}{m_{1}+m_{2}}$ $(i=1,2)$, $m_{1}$, $m_{2}$, and $M$ are the masses of the quark, antiquark, and the initial meson, respectively. $Q_{1}e$ and $Q_{2}e$ are the charges of quark and antiquark, respectively. We use subscript $i$ representing the initial state and $f$ the final state, so $\varphi^{++}_{_{i}}$ and $\varphi^{++}_{_{f}}$ are the positive energy wave functions of the initial and final mesons, respectively.

\subsection{The relativistic wave functions and the mass of $2^{-}$ state}

In this paper, the relativistic wave functions are obtained by solving the instantaneous BS equation, namely the complete Salpeter equation. First, we provide the general representation of the relativistic wave function of a meson according to its quantum number of $J^{P}$ or $J^{PC}$, where the radial wave functions are unknown, $J$, $P$ and $C$ represent total angular momentum, the parity, and charge conjugate parity, respectively. Second, substitute this wave function into the Salpeter equation for solution, and obtain the numerical values of the radial wave functions.
For simplicity, we will not introduce the BS equation and the Salpeter equation here, the interested readers are referred to the original papers \cite{Salpeter:1951sz,Salpeter:1952ib} or our previous articles, for example, Ref. \cite{Kim:2003ny}.

\subsubsection{Mass and wave function of $2^{-}$ state}

The general relativistic wave function for a $2^{-}$ meson in the instantaneous approach can be written as \cite{Wang:2022cxy}
{\begin{eqnarray}\label{2-}
&&\varphi_{_{2^{-}}}(q_{_\bot})=\epsilon_{\mu\nu}q^{\mu}_{_\bot}q^{\nu}_{_\bot}\left(a_{_1}+\frac{\slashed{P}}{M}a_{_2}+\frac{\slashed{q}_{_\bot}}{M}a_{_3}+\frac{\slashed{P}\slashed{q}_{_\bot}}{M^{2}}a_{_4}\right)
\gamma^{5}\nonumber\\&&\hspace{2cm}
+\frac{i\varepsilon_{_{\mu\nu\alpha\beta}}\gamma^{\mu}P^{\nu}q^{\alpha}_{_\bot}\epsilon^{\beta\delta}q_{_{_\bot\delta}}}{M}\left(h_{_{1}}+\frac{\slashed{P}}{M}h_{_2}+\frac{\slashed{q}_{_\bot}}{M}h_{_3}+
\frac{\slashed{P}\slashed{q}_{_\bot}}{M^{2}}h_{_4}\right),
\end{eqnarray}}
where $\epsilon_{\mu\nu}$ is the polarization tensor of the $2^{-}$ state, and $\varepsilon_{_{\mu\nu\alpha\beta}}$ is the Levi-Civita symbol. {$a_{_i}$ and $h_{_i}$ ($i=1,2,3,4$) are the radial wave functions, and they are function of $-{q}^2_{_\bot}$.}
Not all the radial wave functions are independent, only half of them are, and according to the Salpeter equation, we have the relations
$$a_3=-a_1 \frac{M(\omega_{1}-\omega_{2})}{m_1\omega_{2}+m_2\omega_{1}},~a_4=-a_2 \frac{M(\omega_{1}+\omega_{2})}{m_1\omega_{2}+m_2\omega_{1}},$$
{$$h_3=h_1 \frac{M(\omega_{1}-\omega_{2})}{m_1\omega_{2}+m_2\omega_{1}},~h_4=h_2 \frac{M(\omega_{1}+\omega_{2})}{m_1\omega_{2}+m_2\omega_{1}}.$$}
where $\omega_1=\sqrt{m^2_1-{q}^2_{_{\bot}}}$ and $\omega_2=\sqrt{m^2_2-{q}^2_{_{\bot}}}$ are the energies of quark $1$ and antiquark $2$, respectively.

It can be checked that each term in Eq.(\ref{2-}) has a quantum number of $2^-$. With this wave function form as input, we solve the corresponding Salpeter equation and obtain the eigenvalues and the numerical values of the radial wave functions of all the $2^-$ states \cite{Wang:2022cxy}, including the charmonia.
That is to say, when the input masses of quark and antiquark are equal and both are charm quark mass, the system corresponds to charmonia. At this point, the solutions given by the Salpeter equation include both $2^{--}$ and $2^{-+}$ states. The first and second solutions are both ground $1D$ states because the radial wave functions have no nodes. {And the wave function of the first solution contains only non-zero $h_i$ terms ($h_3=0$ for charmonium),} while all $a_i$ terms are zero, so it corresponds to the ${}^3D_{_2}$ $2^{--}$ state. The second solution is opposite to the first one, {as its wave function only contains $a_i$ terms ($a_3=0$ for charmonium) and all $h_i$ terms are zero,} therefore it is the  ${}^1D_{_2}$ $2^{-+}$ state. The third and fourth solutions are similar to the first and second ones, corresponding to $2^{--}$ and $2^{-+}$ states, respectively, but they both have one node in their wave functions, so they are excited $2D$ states, etc.

\begin{figure}[!htb]
\begin{minipage}[c]{1\textwidth}
\includegraphics[width=3in]{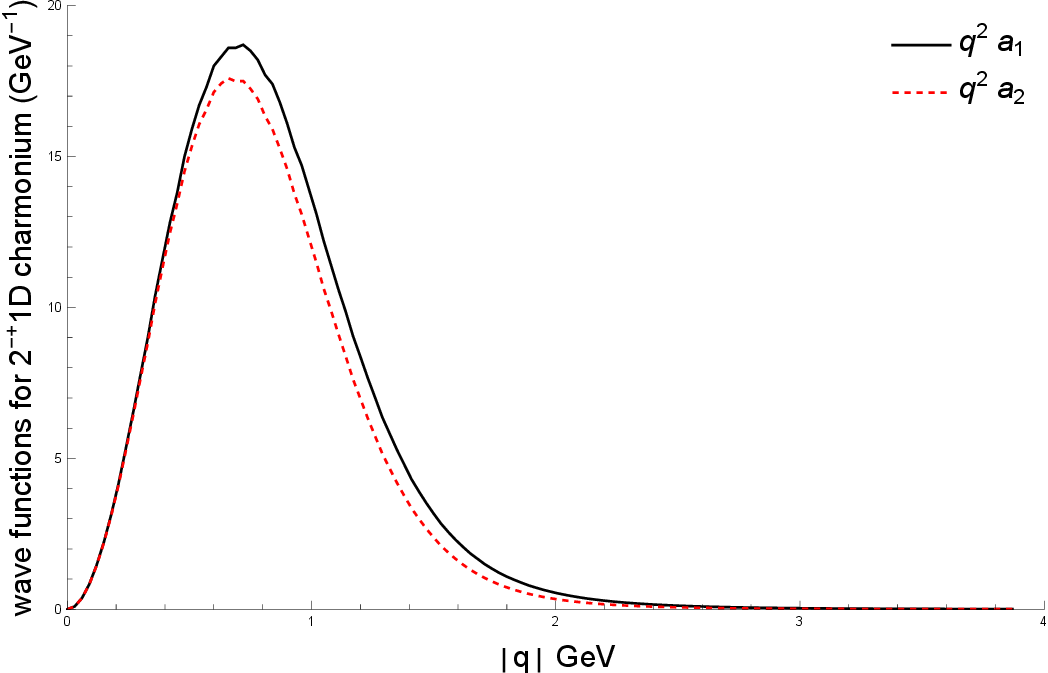}
\includegraphics[width=3in]{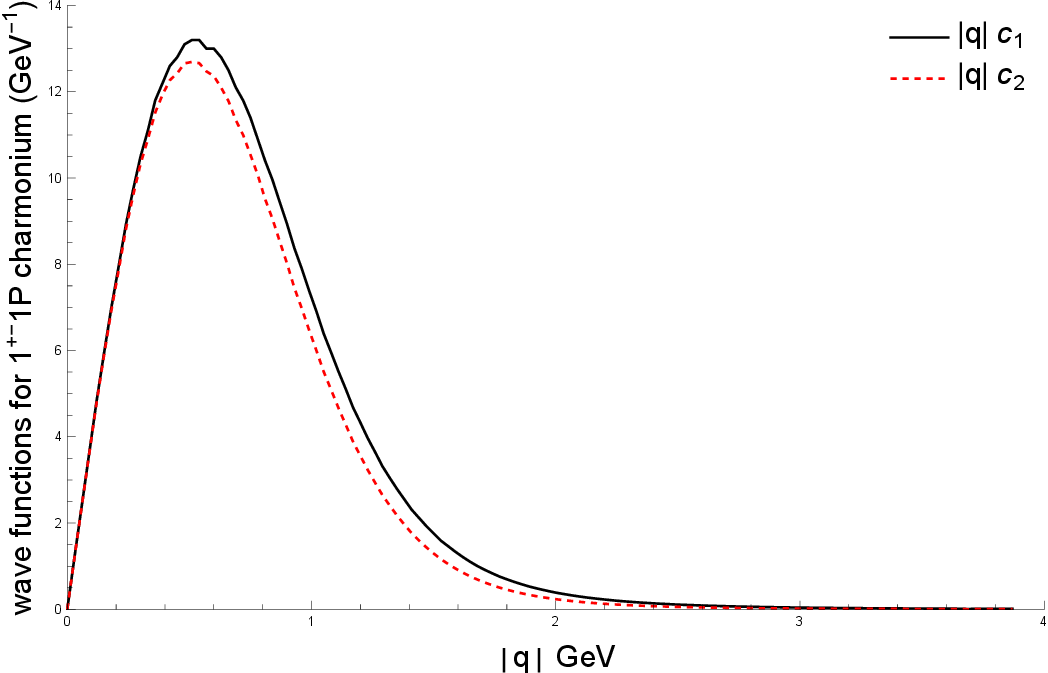}
\end{minipage}
\caption{The radial wave functions of the ground $2^{-+}$ and $1^{+-}$ charmonia.}
\label{wave2-}
\end{figure}

In our calculations, there are some parameters, such as quark mass, etc. To determine these parameters, the mass of the first solution is our input. Fortunately, the mass of the ground $2^{-+}$ state is slightly higher than that of the $2^{--}$ state in our model, and the ground $2^{--}$ state $X(3823)$ has been found in experiment \cite{Belle:2013ewt,BESIII:2015iqd}, so we can predict the ground mass of $2^{-+}$ state successfully. Our predictions of mass spectra of $2^{--}$ and $2^{-+}$ charmonia are shown in Table \ref{mass}, where the ground mass of $2^{--}$ state, 3823.0 MeV, is input, and the predicted mass of ground $2^{-+}$ charmonium is 3828.2 MeV. The diagrams of independent radial wave functions $a_1$ and $a_2$ for the ground $2^{-+}$ charmonium are shown in Figure \ref{wave2-}.

\begin{table*}[hbt]
\caption{Our predictions of the mass spectra of $2^{--}$ and $2^{-+}$ charmonium in unit of MeV}\label{mass}
\begin{tabular*}{\textwidth}{@{}c@{\extracolsep{\fill}}ccc}
\hline \hline
 &$2^{--}$&$2^{-+}$
\\\hline
1D &3823.0~(input)&3828.2
\\
2D &4153.7&4157.7
\\
3D &4407.5&4410.7
\\
\hline\hline
\end{tabular*}
\end{table*}

{For the $2^{-+}$ state, its relativistic wave function is
\begin{eqnarray}\label{2-+1}
&&\varphi_{_{2^{-+}}}(q_{_\bot})=\epsilon_{\mu\nu}q^{\mu}_{_\bot}q^{\nu}_{_\bot}\left(a_{_1}+\frac{\slashed{P}}{M}a_{_2}
+\frac{\slashed{P}\slashed{q}_{_\bot}}{M^{2}}a_{_4}\right)
\gamma^{5},
\end{eqnarray}
and the corresponding positive energy wave function can be expressed as \cite{wth2013}}
\begin{eqnarray}\label{2-+}
\varphi_{_{2^{-+}}}^{++}(q_{_\bot})=\epsilon_{\mu\nu}q^{\mu}_{_\bot}q^{\nu}_{_\bot}\left[A_{_1}+\frac{\slashed{P}}{M}A_{_2}
+\frac{\slashed{P}\slashed{q}_{_\bot}}{M^{2}}A_{_3}\right]\gamma^{5},
\end{eqnarray}
where $A_{_i}$ are related to the original radial wave functions $a_{_i}$,
$$
A_{_1}=\frac{1}{2}\left(a_{_1}+\frac{\omega_{_1}}{m_{_1}}a_{_2}\right),~~
A_{_2}=\frac{m_{_1}}{2\omega_{_1}}\left(a_{_1}+\frac{\omega_{_1}}{m_{_1}}a_{_2}\right),
~~A_{_3}=-\frac{A_{_2}M}{m_{_1}},
$$
where $\omega_{_1}=\omega_{_2}$, and $m_{_1}=m_{_2}$ have been used.

As a relativistic method, the wave functions contain a wealth of information.
First, in the nonrelativistic limit, there are no $a_3$ and $a_4$ terms in Eq.(\ref{2-}), so they are relativistic corrections, while $a_1$ and $a_2$ terms are nonrelativistic. However, in our method, unlike other nonrelativistic methods, they are independent and not equal, that is, $a_1\neq a_2$, see Fig. \ref{wave2-}. Second, it can be seen that the terms $a_1$ and $a_2$ in Eq.(\ref{2-}), or $A_1$ and $A_2$ terms in Eq.(\ref{2-+}) are $D$-wave, but the $a_3$ and $a_4$ terms, or $A_3$ term, are $F$-wave. {To calculate the proportion of $D$ and $F$ waves in the wave function, we show the normalization condition
\begin{equation}
\int\frac{d\vec{q}}{(2\pi)^{3}}\frac{8w_{1}\vec{q}^{4}}{15M m_{1}}a_{1}a_{2}=1.
\end{equation}
In this normalization formula, the contribution of pure $D$ wave is
$$\int\frac{d\vec{q}}{(2\pi)^{3}}\frac{8m_{1}\vec{q}^{4}}{15Mw_{1}}a_{1}a_{2}.$$
Using the upper two formulas, the ratio of $D$-wave and $F$-wave in $2^{-+}$ state $\eta_{c2}$ is calculated as $D: F$= $1:0.151$. This result shows that in the wave function of $2^{-+}$ state, the proportion of nonrelativitic $D$-wave is large and will generally provide main contribution, the proportion of relativistic $F$-wave is small, and it usually play a secondary role.}

\subsubsection{Wave function of $1^{--}$ state}
{The general relativistic wave function for a $1^{-}$ state can be written as \cite{Wang:2005qx}:
\begin{eqnarray}\label{1-}
&&\varphi_{_{1^{-}}}(q_{_{f_{_\bot}}})=\left(\epsilon_{_{f}}\cdot q_{_{f_{_\bot}}}\right)\left[b_{_1}+\frac{\slashed{P}_{_{f}}}{M_{_{f}}}b_{_2}+\frac{\slashed{q}_{_{f_{_\bot}}}}{M_{_{f}}}b_{_3}+\frac{\slashed{P}_{_{f}}\slashed{q}_{_{f_{_\bot}}}}{M^{2}_{_{f}}}b_{_4}\right]
+M_{_{f}}\slashed{\epsilon}_{_{f}}b_{_5}+\slashed{\epsilon}_{_{f}}\slashed{P}_{_{f}}b_{_6}
\nonumber\\&&\hspace{2cm}
+\left(\slashed{q}_{_{f_{_\bot}}}\slashed{\epsilon}_{_{f}}
-\epsilon_{_{f}}\cdot q_{_{f_{_\bot}}}\right)b_{_7}+\frac{1}{M_{_{f}}}\left(\slashed{P}_{_{f}}\slashed{\epsilon}_{_{f}}\slashed{q}_{_{f_{_\bot}}}-\slashed{P}_{_{f}}\epsilon_{_{f}}\cdot q_{_{f_{_\bot}}}\right)b_{_8},
\end{eqnarray}
where $\epsilon_{_{f}}$ is the polarization vector of the meson, $P_{_f}$ and $M_{_f}$ are the total momentum and mass of the $1^-$ state, respectively. There are 8 radial wave functions, while due to the Salpeter equation, only 4 of them are independent, we choose $b_{_3}$, $b_{_4}$, $b_{_5}$ and $b_{_6}$ as the independent radial wave functions. The normalization condition of the $1^{--}$ wave function is as follows:
\begin{equation}\label{1-nor}
\int\frac{d\vec{q}_{_{f}}}{(2\pi)^{3}}\frac{4M_{_{f}}w_{_{f}}}{3m_{_{f}}}\left[-3b_{5}b_{6}+\frac{\vec{q}^{2}_{_{f}}}{M^{2}_{_{f}}}(-b_{4}b_{5}+b_{3}b_{6}+
b_{3}b_{4}\frac{\vec{q}^{2}_{_{f}}}{M^{2}_{_{f}}})\right]=1,
\end{equation}
where $m_{_{f}}=m_{_1}$ and $w_{_{f}}$ are the mass and energy of the quark, respectively.
In the wave function of Eq. (\ref{1-}), the terms including $b_{_{5}}$ and $b_{_{6}}$ are $S$-wave, the $b_{_{1}}, b_{_{2}}$, $b_{7}$ and $b_{8}$ terms are $P$-wave, while $b_{_{3}}$ and $b_{_{4}}$ terms are $D$-wave mixed with $S$-wave, because
\begin{eqnarray}
&&(\epsilon_{_{f}}\cdot q_{_{f_{_\bot}}})\slashed{q}_{_{f_{_\bot}}}=\frac{1}{3}q_{_{f_{_\bot}}}^{2}\slashed{\epsilon}_{_{f}}+\left[(\epsilon_{_{f}}\cdot q_{_{f_{_\bot}}})\slashed{q}_{_{f_{_\bot}}}-\frac{1}{3}q_{_{f_{_\bot}}}^{2}\slashed{\epsilon}_{_{f}}\right],
\end{eqnarray}
where $\frac{1}{3}q_{_{f_{_\bot}}}^{2}\slashed{\epsilon}_{_{f}}$ is $S$-wave and $(\epsilon_{_{f}}\cdot q_{_{f_{_\bot}}})\slashed{q}_{_{f_{_\bot}}}-\frac{1}{3}q_{_{f_{_\bot}}}^{2}\slashed{\epsilon}_{_{f}}$ is $D$-wave. So in Eq. (\ref{1-}), the pure $S$-wave is
\begin{eqnarray}\label{pure1-s}
&&M_{_{f}}\slashed{\epsilon}_{_{f}}\left[b_{_{5}}+\frac{\slashed{P}_{_{f}}}{M_{_{f}}}b_{_{6}}\right]
+\frac{1}{3}q_{_{f_{_\bot}}}^{2}\slashed{\epsilon}_{_{f}}\left[\frac{1}{M_{_{f}}}b_{_{3}}
-\frac{\slashed{P}_{_{f}}}{M^{2}_{_{f}}}b_{_{4}}\right],
\end{eqnarray}
and the pure $D$-wave is
\begin{eqnarray}\label{pure1-d}
&&\left[\epsilon_{_{f}}\cdot q_{_{f_{_\bot}}}\slashed{q}_{_{f_{_\bot}}}-\frac{1}{3}q_{_{f_{_\bot}}}^{2}\slashed{\epsilon}_{_{f}}\right]\left[\frac{1}{M_{_{f}}}b_{_{3}}
-\frac{\slashed{P}_{_{f}}}{M^{2}_{_{f}}}b_{_{4}}\right].
\end{eqnarray}
If only pure $S$-wave is considered, its contribution to the normalization formula, Eq. (\ref{1-nor}), is:
$$-\int\frac{d\vec{q}_{_{f}}}{(2\pi)^{3}}\frac{4m_{_{f}}(3M^{2}_{_{f}}b_{_5}-\vec{q}^{2}_{_f}b_{_3})
(3M^{2}_{_{f}}b_{_6}+\vec{q}^{2}_{_{f}}b_{_4})}{9M^{3}_{_{f}}w_{_{f}}}.$$
And the contribution of pure $D$-wave to the normalization condition is:
$$\int\frac{d\vec{q}_{_{f}}}{(2\pi)^{3}}\frac{8m_{_{f}}\vec{q}^{4}_{_{f}}b_{_3}b_{_4}}{9M^{3}_{_{f}}w_{_{f}}}.$$}
The corresponding positive energy wave function of the $1^{--}$ state is expressed as \cite{Wang:2005qx},
\begin{eqnarray}\label{1--}
\varphi_{_{1^{--}}}^{++}(q_{_{f_{_\bot}}})=&&\left(\epsilon_{_{f}}\cdot q_{_{f_{_\bot}}}\right)\left[B_{_1}+\frac{\slashed{P}_{_{f}}}{M_{_{f}}}B_{_2}+\frac{\slashed{q}_{_{f_{_\bot}}}}{M_{_{f}}}B_{_3}+\frac{\slashed{P}_{_{f}}\slashed{q}_{_{f_{_\bot}}}}{M^{2}_{_{f}}}B_{_4}\right]
\nonumber\\
&&+M_{_{f}}\slashed{\epsilon}_{_{f}}\left[B_{_5}+\frac{\slashed{P}_{_{f}}}{M_{_{f}}}B_{_6}+\frac{\slashed{P}_{_{f}}\slashed{q}_{_{f_{_\bot}}}}{M^{2}_{_{f}}}B_{_7}\right],
\end{eqnarray}
where $B_{_i}(i=1,2,\cdots,7)$ is a function of the four independent radial wave functions $b_{_3}, b_{_4}, b_{_5}$ and $b_{_6}$ of the $1^{--}$ state, whose numerical values are obtained by solving the corresponding Salpeter equation \cite{Wang:2005qx}, and we have,
$$f=\frac{1}{2}\left(b_{_{3}}+\frac{m_{_{f}}}{w_{_{f}}}b_{_{4}}\right),~~
B_{_{1}}=\frac{q_{_{f_{_\bot}}}^{2}}{M_{_{f}}m_{_{f}}}f+\frac{M_{_{f}}}{2m_{_{f}}}\left(b_{_5}-\frac{m_{_{f}}}{w_{_{f}}}b_{_6}\right),~~
B_{_5}=\frac{1}{2}\left(b_{_5}-\frac{w_{_{f}}}{m_{_{f}}}b_{_6}\right),~~$$
$$B_{_2}=-\frac{M_{_{f}}}{w_{_{f}}}B_{_5},~~
B_{_3}=f-\frac{M^{2}_{_{f}}}{2m_{_{f}}w_{_{f}}}b_{_6},~~
B_{_4}=\frac{w_{_{f}}}{m_{_{f}}}f-\frac{M^{2}_{_{f}}}{2m_{_{f}}w_{_{f}}}b_{_5},~~
B_{_6}=-\frac{m_{_{f}}}{w_{_{f}}}B_{_5},~~
B_{_7}=B_{_2},$$

\begin{figure}[!htb]
\begin{minipage}[c]{1\textwidth}
\includegraphics[width=3in]{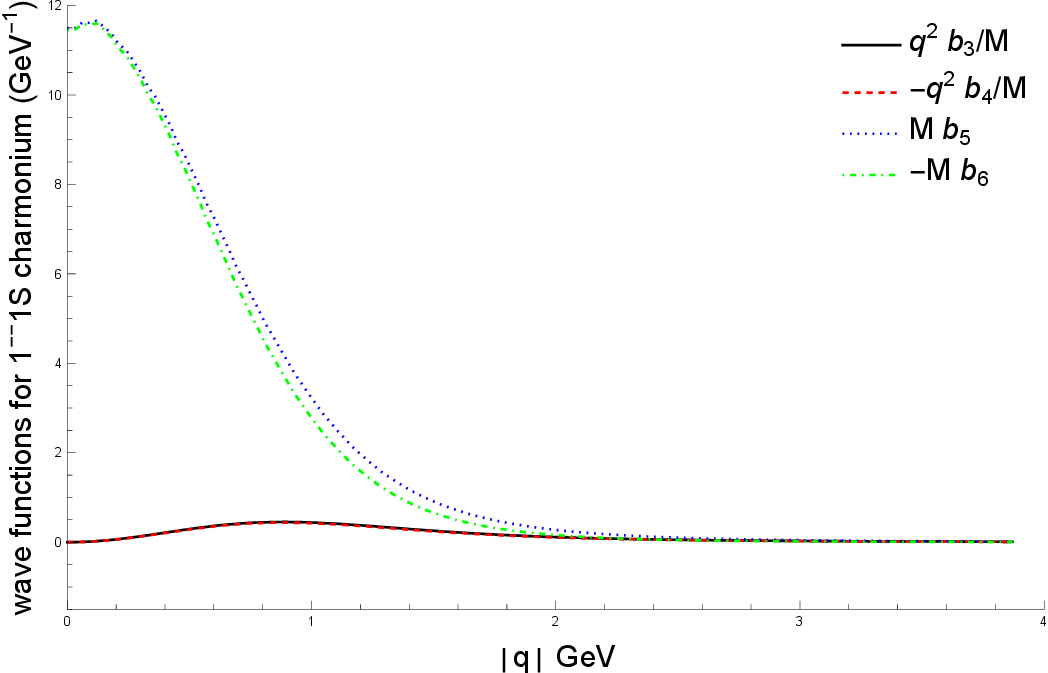}
\includegraphics[width=3in]{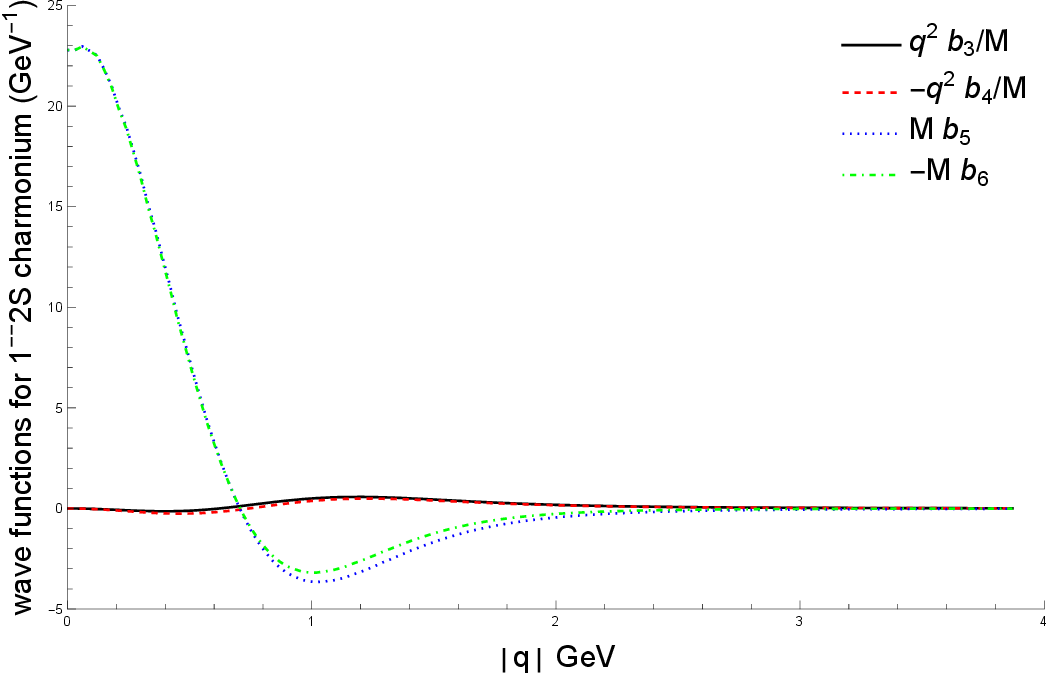}
\includegraphics[width=3in]{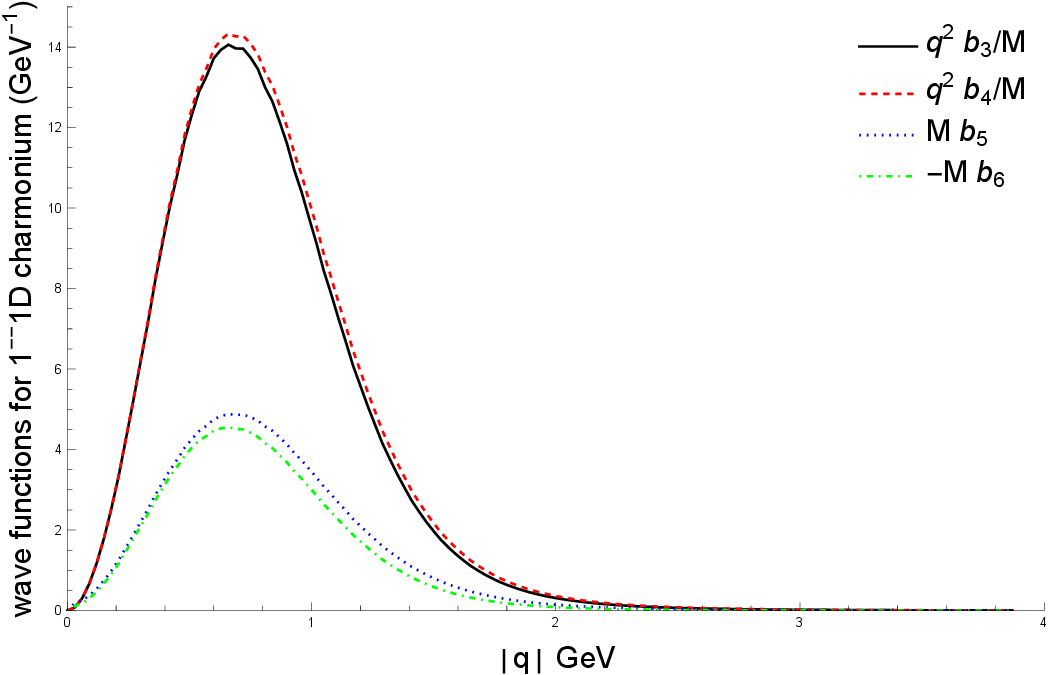}
\end{minipage}
\caption{The radial wave functions for the ground, first excited and second excited $1^{--}$ states.}
\label{wave1-}
\end{figure}

Since the expression of Eq.(\ref{1--}) is a general relativistic form for the wave function of $1^{--}$ state, with this as input, as a relativistic dynamic equation for bound state, the solution of the Salpeter equation includes all possible $1^{--}$ states, including states dominated by $S$-waves and states dominated by $D$-waves. {And all the obtained states are $S-P-D$ mixed states. We show the diagrams of the independent radial wave functions of the first three solutions in Figure \ref{wave1-}. It can be seen that the wave functions of the ground state and the first excited state are dominant by $S$-waves, and their eigenvalues are $3096.9$ MeV and $3688.1$ MeV, so they are $J/\psi$ and $\psi(2S)$, respectively. We also calculate the proportion of different partial waves in the wave function, and obtain $S: P: D= 1:0.126:0.0551$ and $S: P: D= 1:0.148:0.0647$ for $J/\psi$ and $\psi(2S)$, respectively. The second excited state is dominated by $D$-waves and there are no nodes in the wave functions, and its predicted mass is $3778.9$ MeV, so it is the particle of $\psi(3770)$ \cite{Chang:2010kj}. The corresponding ratio of $S: P: D= 0.0631:0.171:1$ is obtained for $\psi(3770)$ in our method.}

\subsubsection{Wave function of $1^{+-}$ state}

{The general relativistic wave function for the $1^{+-}$ state can be written as \cite{Wang:2012pf}:
\begin{eqnarray}\label{1+-}
&&\varphi_{_{1^{+-}}}(q_{_{f_{_\bot}}})=\epsilon_{_{f}}\cdot q_{_{f_{_\bot}}}\left(c_{_1}+c_{_2}\frac{\slashed{P}_{_{f}}}{M_{_{f}}}+c_{_3}\frac{\slashed{q}_{_{f_{_\bot}}}}{M_{_{f}}}+c_{_4}\frac{\slashed{P}_{_{f}}\slashed{q}_{_{f_{_\bot}}}}{M^{2}_{f}}\right)\gamma^{5},
\end{eqnarray}
where $\epsilon_{_{f}}$ is the polarization vector of the $1^{+-}$ state; $c_{_1}$ and $c_{_2}$ are the independent radial wave functions, and $c_{_3}$ and $c_{_4}$ are functions of them. The normalization condition of this wave function is:
\begin{equation}\int\frac{d\vec{q}_{_{f}}}{(2\pi)^{3}}\frac{4w_{_{f}}
\vec{q}^{2}_{_{f}}}{3M_{_{f}}m_{_{f}}}c_{_1}c_{_2}=1.\end{equation}
In Eq.(\ref{1+-}), we note that the terms with $c_{_1}$ and $c_{_2}$ are $P$ waves, and they give nonrelativistic contribution; $c_{_3}$ and $c_{_4}$ term are $D$ waves, and they provide the relativistic corrections. In the formula of normalization condition, the pure $P$-wave has the following contribution:
$$\int\frac{d\vec{q}_{_{f}}}{(2\pi)^{3}}\frac{4m_{_{f}}\vec{q}^{2}_{_{f}}}{3M_{_{f}}w_{_{f}}}c_{_1}c_{_2}.$$}

The corresponding positive energy wave function of the $1^{+-}$ state can be expressed as \cite{Wang:2012pf},
\begin{eqnarray}\label{1+-p}
&&\varphi_{_{1^{+-}}}^{++}(q_{_{f_{_\bot}}})=\left(\epsilon_{_{f}}\cdot q_{_{f_{_\bot}}}\right)\left[C_{_{1}}+\frac{\slashed{P}_{_{f}}}{M_{_{f}}}C_{_{2}}+
\frac{\slashed{P}_{_{f}}\slashed{q}_{_{f_{_\bot}}}}{M^{2}_{f}}C_{_{3}}\right]\gamma^{5},
\end{eqnarray}
where the $C_{_i}$ are related to the original radial wave functions $c_{_i}$
$$
C_{_{1}}=\frac{1}{2}\left(c_{_{1}}+\frac{w_{_{f}}}{m_{_{f}}}c_{_{2}}\right),~~
C_{_{2}}=\frac{1}{2}\left(\frac{m_{_{f}}}{w_{_{f}}}c_{_{1}}+c_{_{2}}\right),~~
C_{_{3}}=-\frac{M_{f}}{w_{_{f}}}C_{_{1}},$$
and we show the diagrams of independent radial wave functions $c_{_{1}}$ and $c_{_{2}}$ in the right part of Fig.\ref{wave2-}. The mass of ground $1^{+-}$ charmonium $h_c(1P)$ is 3525 MeV.
{For $h_c(1P)$, we obtain the ratio $P: D$= 1: 0.131. So $P$-wave is dominant and it gives the nonrelativistic contribution, and the wave function of $h_c(1P)$ also contains a small component of $D$-wave providing the relativistic correction.}

\subsection{The form factors}

For $\eta_{_{c2}}\rightarrow\psi\gamma$, we substitute Eq.(\ref{2-+}) and Eq.(\ref{1--}) into Eq.(\ref{em}), after integrate the internal momentum $q_{_{\perp}}$ over the initial and final state wave functions, we obtain the EM transition matrix element represented by the form factors,
\begin{eqnarray}
&&{\cal M}^{\xi}_{_{2^{-+}\rightarrow1^{--}}}=2\varepsilon^{{\xi PP_{_{f}}\nu}}\epsilon_{_{\nu P_{_{f}}}}P\cdot\epsilon_{_{f}}t_{_{1}}
+2\varepsilon^{{\xi PP_{_{f}}\nu}}\epsilon_{_{\nu\epsilon_{_{f}}}}t_{_{2}}+2\varepsilon^{{\xi \epsilon_{f}P\nu}}\epsilon_{_{\nu P_{_{f}}}}t_{_{3}}
\nonumber\\&&\hspace{2.4cm}
+2\varepsilon^{{\xi \epsilon_{f}P_{f}\nu}}\epsilon_{_{\nu P_{_{f}}}}t_{_{4}}+\varepsilon^{{\xi \epsilon_{f}PP_{f}}}\epsilon_{_{P_{_{f}}P_{_{f}}}}t_{_{5}}+
2P^{\xi}\varepsilon^{{\epsilon_{f}PP_{f}\nu}}\epsilon_{_{\nu P_{_{f}}}}t_{_{6}},
\end{eqnarray}
where $t_{i}$ ($i=1,2...6$) is the form factor. And we have used some abbreviations here, for examples, $\varepsilon^{{\xi PP_{_{f}}\nu}}\equiv\varepsilon^{\xi\alpha\beta\nu}P_{\alpha}{P_{_f}}_{\beta}$
 and $\epsilon_{_{P_{_{f}}P_{_{f}}}}=\epsilon_{\alpha\beta}{P^{\alpha}_{_{f}}P_{_{f}}^{\beta}}$.

Similarly, for $\eta_{_{c2}}\rightarrow h_{_{c}}\gamma$, we insert Eq.(\ref{2-+}) and Eq.(\ref{1+-}) into Eq.(\ref{em}), and obtain
\begin{eqnarray}\label{Mhc}
&&{\cal M}^{\xi}_{_{2^{-+}\rightarrow1^{+-}}}=P^{\xi}\epsilon_{_{P_{_{f}}P_{_{f}}}}P\cdot\epsilon_{_{f}}s_{_1}
+P^{\xi}_{_{f}}\epsilon_{_{P_{_{f}}P_{_{f}}}}P\cdot\epsilon_{_{f}}s_{_2}+2\epsilon^{\xi P_{_{f}}}P\cdot\epsilon_{_{f}}s_{_3}
\nonumber\\&&\hspace{2.3cm}
+2P^{\xi}\epsilon_{_{P_{_{f}}} \epsilon_{_{f}}}s_{_4}+2P^{\xi}_{_{f}}\epsilon_{_{P_{_{f}}} \epsilon_{_{f}}}s_{_5}
+\epsilon^{\xi}_{_{f}}\epsilon_{_{P_{_{f}}P_{_{f}}}}s_{_6}
+2\epsilon^{\xi \epsilon_{_{f}}}s_{_{7}},
\end{eqnarray}
where $s_{i}$ ($i=1,2...7$) is the form factor. Since the expressions of form factors $t_{i}$ and $s_{i}$ are very complex, we will not show the details.

For EM transition, we note that not all form factors are independent. Due to the gauge invariant condition $(P_{_\xi}-P_{_{f}\xi}){\cal M}^{\xi}=0$, they satisfy the following constraints,
\begin{eqnarray}
&&t_{_3}=(M^2-ME_{_f})t_{_6}-t_{_4},
\nonumber\\&&s_{_6}=(ME_{_f}-M^2)s_{_1}-(ME_{_f}-M^2_{f})s_{_2}+2s_{_3},
\nonumber\\&&s_{_7}=(M^2-ME_{_f})s_{_4}+(ME_{_f}-M^2_{f})s_{_5}.
\end{eqnarray}


\section{NUMERICAL RESULTS AND DISCUSSIONS}

In our calculation, the charm quark mass is chosen as $m_{c}=1.62$ GeV, and all the meson masses used have been mentioned in the text, for example, the predicted mass of $\eta_{_{c2}}(^1D_{_2})$ is $3828.2$ MeV.

\subsection{EM decay width of the $\eta_{_{c2}}(^1D_{_2})$}

As a $2^{-+}$ state, the main EM decay of $\eta_{_{c2}}(^1D_{_2})$ is $\eta_{_{c2}}(^1D_{_2})\to h_{_{c}}(1P)\gamma$, which is mainly an $E1$ transition. In our calculation, due to the use of relativistic wave functions, our result includes not only the nonrelativistic $E1$ transition, but also the relativistic $M2$ and $E3$ transitions \cite{liwei1}, see Sec.III.B. Our result is
\begin{eqnarray}
&&\Gamma\left[\eta_{_{c2}}(1^1D_{_2})\rightarrow h_{_{c}}(1P)\gamma\right]=284~\textrm{keV}.
\end{eqnarray}
The $\eta_{_{c2}}(^1D_{_2})$ can also decay to a $1^{--}$ charmonium, which is a $M1$ transition in a nonrelativistic method, but is $M1+E2+M3+...$ in our relativistic method. There are three channels of such decays, and our predictions are
\begin{eqnarray}
&&\Gamma\left[\eta_{_{c2}}(1^1D_{_2})\rightarrow J/\psi\gamma\right]=1.04~\textrm{keV},
\end{eqnarray}
\begin{eqnarray}
&&\Gamma\left[\eta_{_{c2}}(1^1D_{_2})\rightarrow\psi(2S)\gamma\right]=3.08~\textrm{eV},
\end{eqnarray}
\begin{eqnarray}
&&\Gamma\left[\eta_{_{c2}}(1^1D_{_2})\rightarrow\psi(3770)\gamma\right]=0.143~\textrm{keV}.
\end{eqnarray}
The results show that the decay width of $\eta_{_{c2}}(1^1D_{_2})\rightarrow h_{_{c}}(1P)\gamma$ is more than two orders of magnitude wider than those of other channels.

\begin{table}[htb]
\caption{The mass of particle $\eta_{_{c2}}(1^1D_{_2})$ and its EM decay widths, as well as the ratios of decay widths.}
\resizebox{\textwidth}{!}
{
\begin{tabular}{c|cccccccc}
\hline
& ~~ours ~~ &  ~~ {\cite{Sebastian:1996cy}}~~&  ~~{\cite{Eichten:2002qv}}~~&~~\cite{Ebert:2002pp}~~ & ~~{\cite{Barnes:2005pb}}~~ &~~{\cite{Li:2009zu}}~~  & ~~{\cite{Jia:2010jn}}~~& ~~{\cite{Deng:2016stx}} ~~ \\
\hline
$M_{(\eta_{_{c2}}(^1D_{_2}))}($\rm MeV$)$ & $3828.2$ & $3820$& $3825$&3811 & $3837(3872)$ & $3796$   & $3872$& $3820$  \\
\hline
$\Gamma(\eta_{_{c2}}(1^1D_{_2})\rightarrow h_{_{c}}(1P)\gamma)($\rm keV$)$ & 284 & 288& 303&245& 344(464) & 375   & $587\sim786$ & 362  \\
\hline
$\Gamma(\eta_{_{c2}}(1^1D_{_2})\rightarrow J/\psi\gamma)($\rm keV$)$ & 1.04  & 0.699  & & &  &   & $3.11\sim4.78$& \\
 \hline
$\Gamma(\eta_{_{c2}}(1^1D_{_2})\rightarrow\psi(2S)\gamma)($\rm eV$)$ & 3.08 & 1&  &  &  & & $17\sim29$&   \\
\hline
$\Gamma(\eta_{_{c2}}(1^1D_{_2})\rightarrow\psi(3770)\gamma)($\rm keV$)$ & 0.143 &  &   0.34 & & &  & $0.49\sim0.56$ & \\
 \hline
$\frac{\Gamma(\eta_{_{c2}}(1^1D_{_2})\rightarrow J/\psi\gamma)}
{\Gamma(\eta_{_{c2}}(1^1D_{_2})\rightarrow h_{_{c}}(1P)\gamma)}$$\times10^{-3}$ & $3.66$ & $2.43$&  &  & &   &  $3.96\sim8.14$ & \\
\hline
$\frac{\Gamma(\eta_{_{c2}}(1^1D_{_2})\rightarrow\psi(2S)\gamma)}
{\Gamma(\eta_{_{c2}}(1^1D_{_2})\rightarrow h_{_{c}}(1P)\gamma)}$$\times10^{-5}$ & $1.08$  & $0.347$&  &  &  &  & $2.16\sim4.94$ &  \\
\hline
$\frac{\Gamma(\eta_{_{c2}}(1^1D_{_2})\rightarrow\psi(3770)\gamma)}
{\Gamma(\eta_{_{c2}}(1^1D_{_2})\rightarrow h_{_{c}}(1P)\gamma)}$$\times10^{-4}$ & $5.04$ &  & $11.2$& & &  & $6.23\sim9.54$ &    \\
\hline
\end{tabular}}
\label{I}
\end{table}

For comparison, we present our results and those from other theoretical predictions in Table \ref{I}.
It can be seen from Table \ref{I}, the predicted masses of $\eta_{_{c2}}(^1D_{_2})$ by different models are mostly concentrated in a small range, $3796\sim3837$ MeV. Some theories have attempted the possibility of a mass of 3872 MeV, mainly because they examined the possibility of the new particle $X(3872)$ as the $\eta_{_{c2}}(^1D_{_2})$.
Our predicted mass of $\eta_{_{c2}}(^1D_{_2})$, $3828.2$ MeV, is very close to the masses predicted in Refs. \cite{Eichten:2002qv,Sebastian:1996cy,Deng:2016stx}, which are $3825$ MeV and $3820$ MeV, respectively.

As a electric dipole $E1$ dominant decay, the radiative transition $\eta_{_{c2}}(^1D_{_2}) \to h_{_{c}}(1P)\gamma$ is the dominant decay channel of $\eta_{_{c2}}(^1D_{_2})$. Other decays dominated by magnetic dipole $M1$ transition have very small partial widths. We also show the ratios of $\frac{\Gamma(\eta_{_{c2}}\rightarrow\psi\gamma)}{\Gamma(\eta_{_{c2}}\rightarrow h_{_{c}}\gamma)}$ in Table \ref{I}. It can be seen that the predicted widths of $\eta_{_{c2}}(^1D_{_2}) \to h_{_{c}}(1P)\gamma$ from different models are comparable, except for Ref.\cite{Jia:2010jn} which has a much larger width, which may be due to the use of a heavier mass of $\eta_{_{c2}}(^1D_{_2})$. For the decay processes dominated by $M1$ transition, $\eta_{_{c2}}\rightarrow\psi\gamma$, there are significant differences between theoretical results. We believe that the main reason for this difference is the relativistic correction, which means the electric quadrupole $E2$ transition has a significant contribution to such processes, see the next subsection.

\subsection{Contribution of different partial waves in EM decay}

In our method \cite{Wang:2022cxy}, the construction of the wave function is not based on the wave (i.e. the orbital angular momentum), but on the $J^{P}$ quantum number of the meson. Therefore, the wave function of this meson is not a pure wave, but contains at least two types of partial waves. Among them, what people are familiar with in literature is the main wave of this particle, which generally provides the nonrelativistic result. In addition, other partial waves generally contribute to the relativistic corrections.

In this subsection, we investigate carefully the contributions of different partial waves of initial and final mesons in the EM decay.
The results are shown in Tables \ref{charm1+-1P}, \ref{charm1--1S}, \ref{charm1--2S}, and \ref{charm1--1D}, where $``2^{-+}"$ represents the wave function of initial meson $\eta_{_{c2}}$, $``1^{--}"$ and $``1^{+-}"$ represent the wave functions of final state meson $\psi$ and $h_{_{c}}$, respectively. $``complete"$ means the complete wave function, and $``S~wave"$, for example, represents only the $S$ partial wave contribute, other partial waves are ignored. To distinguish between the initial and final states, we use the symbol $``prime"$ to represent the wave of the final state. So $``D\times S' "$ means that the contribution to the decay width from the interaction between the $D$-wave in the initial state and the $S$-wave in the final state.

\subsubsection{$\eta_{_{c2}}(1^1D_{_2})\rightarrow h_{_{c}}(1P)\gamma$}
Table \ref{charm1+-1P} shows some results of different partial waves contribute to the decay width of $\eta_{_{c2}}(1^1D_{_2})\rightarrow h_{_{c}}(1P)\gamma$ (some contribution of cross terms, for example, $(D\times P')(D\times D')$ are not listed). As can be seen from Table \ref{charm1+-1P}, the $D$-wave in $\eta_{_{c2}}(1^1D_{_2})$ and $P'$ wave in $h_{_{c}}(1P)$ are their respective dominant waves, and $D\times P'$ provide the main nonrelativistic contribution, $266$ keV, to the decay width, which corresponds to the $E1$ decay. $F\times P'$ and $D\times D'$ are the $M2$ decay, their widths of $1.11$ keV and $0.509$ keV are much smaller than those of $E1$ decay. $F\times D'$ is the $E3$ transition, its contribution of $2.29$ keV is a little larger than that of $M2$ decay.
\begin{table}[H]
\begin{center}
\caption{Contribution of different partial waves to the decay width (keV) of $\eta_{_{c2}}(1^1D_{_2})\rightarrow h_{_{c}}(1P)\gamma$.}\label{charm1+-1P}
{\begin{tabular}{|c|c|c|c|c|}
\hline
\diagbox {$2^{-+}$}{$1^{+-}$}     & $complete'$                     &~~~~~~~~~$P'~wave$~~~~~~~~~~   &~~~~~~~~~$D'~wave$~~~~~~~~~~     \\ \hline
  ~~~~~~~~~$complete$~~~~~~~~~    & 284                             & 240                           & 2.28                    \\ \hline
  $D~wave~$                       & ~~~~~~~~~~~~~259~~~~~~~~~~~~~   & 266                           & ~~~~~~~~0.509~~~~~~~~   \\ \hline
  $F~wave~$                       & 1.01                            & 1.11                          & 2.29                    \\ \hline
\end{tabular}}
\end{center}
\end{table}

\subsubsection{$\eta_{_{c2}}(1^1D_{_2})\rightarrow J/\psi\gamma$}

\begin{table}[H]
\begin{center}
\caption{Contribution of different partial waves to the decay width (keV) of $\eta_{_{c2}}(1^1D_{_2})\rightarrow J/\psi\gamma$.}\label{charm1--1S}
{\begin{tabular}{|c|c|c|c|c|}
\hline
\diagbox {$2^{-+}$}{$1^{--}$}     & $complete'$               & $S'~wave$               & $P'~wave$               & $D'~wave$     \\ \hline
  ~~~~~~~$complete$~~~~~~~        & 1.04                      & 0.379                   & 1.65                    & 0.365        \\ \hline
  $D~wave~$                       & ~~~~~~~~~2.81~~~~~~~~~    & ~~~~~~~~0.284~~~~~~~~   & ~~~~~~~~~1.65~~~~~~~~~  &~~~~~~~~0.112~~~~~~~~   \\ \hline
  $F~wave~$                       & 1.15                      & 0.708                   & 0                       & 0.0725       \\ \hline
\end{tabular}}
\end{center}
\end{table}
Table \ref{charm1--1S} shows the results of different partial waves contribute to the decay width of $\eta_{_{c2}}(1^1D_{_2})\rightarrow J/\psi\gamma$. The lowest order contribution of this process comes from the $M1$ transition, which in our method is the nonrelativistic $D\times S'$ with a small value of $0.284$ keV. The second order contribution, which is also the main relativistic correction, comes from the $E2$ transition $D\times P'$ and $F\times S'$, which contribute $1.65$ keV and $0.708$ keV to the decay width, respectively, significantly greater than that of the lowest order. Other relativistic corrections, the $M3$ transition $D\times D'$ provides a partial decay width of $0.112$ keV and $F\times P'=0$ has no contribution, while the $E4$ transition $F\times D'$ gives a width of $0.0725$ keV. So for the decay $\eta_{_{c2}}(1^1D_{_2})\rightarrow J/\psi\gamma$, our relativistic method includes the contributions from transitions such as $M1+E2+M3+E4$, with the largest contribution coming from the $E2$ transition.

{In Table \ref{I}, our relativistic result of $\eta_{_{c2}}(1D)\rightarrow J/\psi\gamma$, is $1.04$ keV, slightly higher than the $0.699$ keV predicted in Ref.\cite{Sebastian:1996cy}, and much smaller than the result of $3.11\sim4.78$ keV in Ref.\cite{Jia:2010jn}. In Table \ref{charm1--1S}, our nonrelativistic result is $0.284$ keV, which is much smaller than both of them. In our calculations, we only normalized the complete wave functions, as shown in Eq.(6) and Eq.(8). To compare the nonrelativistic results, we should normalize the nonrelativistic wave functions in our method separately. After doing so, our nonrelativistic result change from $0.284$ keV to $0.837$ keV, which is very close to the $0.699$ keV predicted by Ref.\cite{Sebastian:1996cy}.}

\subsubsection{$\eta_{_{c2}}(1^1D_{_2})\rightarrow\psi(2S)\gamma$}
\begin{table}[H]
\begin{center}
\caption{Contribution of different partial waves to the decay width (eV) of $\eta_{_{c2}}(1^1D_{_2})\rightarrow\psi(2S)\gamma$.}\label{charm1--2S}
{\begin{tabular}{|c|c|c|c|c|}
\hline
\diagbox {$2^{-+}$}{$1^{--}$}     & $complete'$              & $S'~wave$                &$P'~wave$               &$D'~wave$     \\ \hline
  ~~~~~~~$complete$~~~~~~         & 3.08                     & 0.0401                   & 0.223                  & 2.49        \\ \hline
  $D~wave~$                       & ~~~~~~~~~1.67~~~~~~~~~   & ~~~~~~~~0.0426~~~~~~~~   & ~~~~~~~~0.223~~~~~~~~  &~~~~~~~~0.697~~~~~~~~  \\ \hline
  $F~wave~$                       & 0.367                    & 0.0892                   & 0                      & 0.552       \\ \hline
\end{tabular}}
\end{center}
\end{table}
Table \ref{charm1--2S} shows the results of  $\eta_{_{c2}}(1^1D_{_2})\rightarrow \psi(2S)\gamma$, where we can see that the decay width is very small, in the order of eV. The main reason is that there is a node structure in the wave function of the $2S$ state, see the second diagram in Fig. \ref{wave1-}, and the contributions of the wave functions before and after the nodes cancel to each other, resulting in a very small width. Meanwhile, due to the node structure, it is not easy to draw clear conclusion like we did in $\eta_{_{c2}}(1^1D_{_2})\rightarrow J/\psi\gamma$.
It can only be concluded that the contribution of decay $\eta_{_{c2}}(1^1D_{_2})\rightarrow \psi(2S)\gamma$ mainly comes from relativistic corrections.

\subsubsection{$\eta_{_{c2}}(1^1D_{_2})\rightarrow\psi(3770)\gamma$}
In Table \ref{charm1--1D}, we show the contributions of different partial waves to the decay width of $\eta_{_{c2}}(1^1D_{_2})\rightarrow\psi(3770)\gamma$. As can be seen from Table \ref{charm1--1D}, the $D'$-wave in the final state $\psi(3770)$ provides the dominant contribution, and contributions of $P'$-wave and $S'$-wave can be ignored. For $\eta_{_{c2}}(1^1D_{_2})$, the contributions of $D$-wave and $F$-wave are almost equal.
In Table \ref{charm1--1D}, the lowest order contribution comes from the nonrelativistic $M1$ transition, $D\times D'$, with a partial width of $0.0354$ keV. The second order contribution is the $E2$ transitions $F\times D'$ and $D\times P'$, and the contribution of the former is $0.0359$ keV, which is similar to the contribution of the lowest order.

\begin{table}[H]
\begin{center}
\caption{Contribution of different partial waves to the decay width (keV) of $\eta_{_{c2}}(1^1D_{_2})\rightarrow\psi(3770)\gamma$.}\label{charm1--1D}
{\begin{tabular}{|c|c|c|c|c|}
\hline
\diagbox {$2^{-+}$}{$1^{--}$}     & $complete'$              & $D'~wave$                &$P'~wave$                    &$S'~wave$        \\ \hline
  ~~~~~~~$complete$~~~~~~         & 0.143                    & 0.142                    &$1.12\times10^{-6}$          &$2.19\times10^{-6}$        \\ \hline
  $D~wave~$                       & ~~~~~~~~0.0356~~~~~~~~   & ~~~~~~~~0.0354~~~~~~~~   &~~~~$1.12\times10^{-6}$~~~~  & ~~~~$2.42\times10^{-8}$~~~~  \\ \hline
  $F~wave~$                       & 0.0359                   & 0.0359                   & 0                           &$2.55\times10^{-6}$       \\ \hline
\end{tabular}}
\end{center}
\end{table}

\subsection{Discussion on the full width of $\eta_{_{c2}}(1D)$}
The $\eta_{_{c2}}(1D)$ has not been detected in experiment,
so we like to estimate the full width of $\eta_{_{c2}}$. In addition to EM decay $\eta_{_{c2}}(1D)\rightarrow h_{_{c}}(1P)\gamma$, strong decays of $\eta_{_{c2}}(1D)$ are also important decay channels, which can provide sizable width. Since there is no strong decay allowed by the OZI-rule, the main strong decays of $\eta_{_{c2}}(1D)$ are $\eta_{_{c2}}(1D)\to\eta_c \pi\pi$ \cite{Eichten:2002qv} and $\eta_{_{c2}}(1D)\to gg$ \cite{wth2013}.
{By the approximation that all the $\psi(^3D_{_J})\to J/\psi \pi\pi$ rates and the rate of $\eta_c(^1D_{_2})\to\eta_c \pi\pi$ are equal \cite{yantm,barnes}, we follow the method in Ref. \cite{wangbo}, and obtain
\begin{equation}\Gamma(\eta_{_{c2}}(1D)\to\eta_c \pi\pi)= 144~\rm{keV}.\end{equation}This value is consistent with the result $210\pm110$ keV in Ref. \cite{barnes}. Moreover, as shown in the second graph of Figure \ref{mass1} below, if we extend the result curve to mass of $3770$ MeV, our result is in good agreement with the experimental data $\Gamma(\Psi(3770)\to J/\psi \pi\pi)=74.3\pm 18.5~\rm{keV}$.
In previous paper \cite{wth2013}, we estimated
\begin{equation}\Gamma(\eta_{_{c2}}(1D)\to gg)= 46.1~\rm{keV}.\end{equation}
Using these two sets of values, combined with the EM results, we estimate the total width of $\eta_{_{c2}}(1D)$ to be
\begin{eqnarray}\label{width}
\Gamma(\eta_{_{c2}}(1D))\approx \Gamma( h_{_{c}}\gamma)+\Gamma(\eta_c \pi\pi)+\Gamma(gg)+\Gamma(\psi\gamma)=475~\rm{keV}.
\end{eqnarray}}
{Since neither $\eta_{_{c2}}$ nor its mass has been discovered by experiment, and the large results of Refs.\cite{Barnes:2005pb,Jia:2010jn} in Table \ref{I} may indicate that the partial widths of EM decays are sensitive to the mass of $\eta_{_{c2}}(1D)$. So when estimating the total width of $\eta_{_{c2}}(1D)$, this influence should be taken into account. Therefore, we vary the mass of $\eta_{_{c2}}(1D)$ from 3800 MeV to 3872 MeV, which is a common theoretical predicted mass range about $\eta_{_{c2}}(1D)$. And calculate the corresponding partial widths of the EM decays, strong decays, as well as the total width, and show the results in Figure \ref{mass1}.}
\begin{figure}[!htb]
\begin{minipage}[c]{1\textwidth}
\includegraphics[scale=0.4]{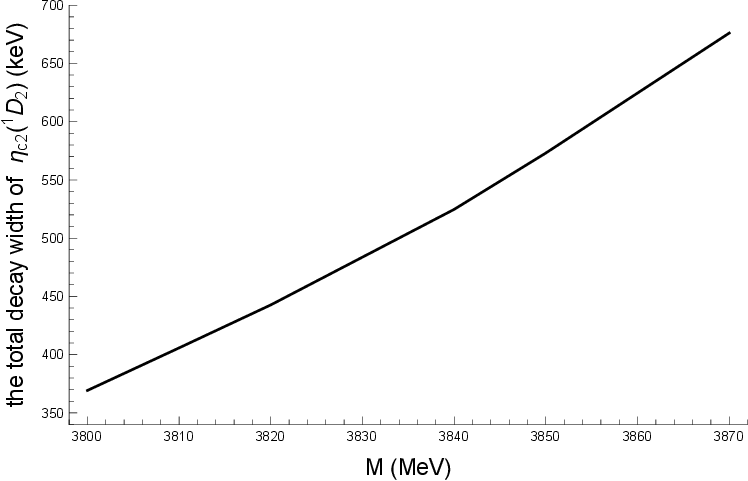}
    \includegraphics[scale=0.4]{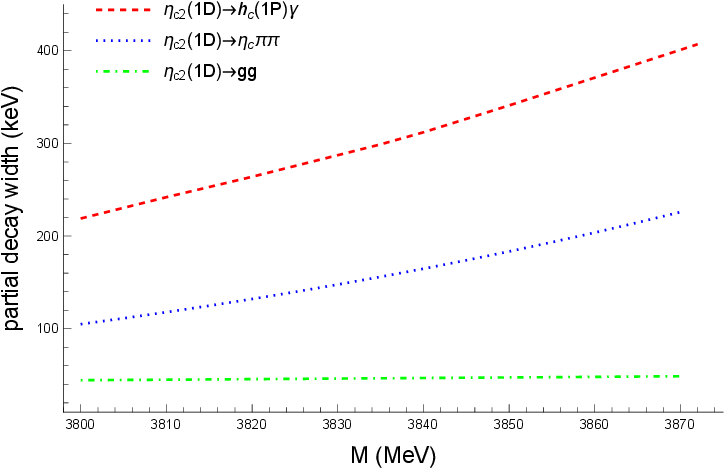}
    \includegraphics[scale=0.4]{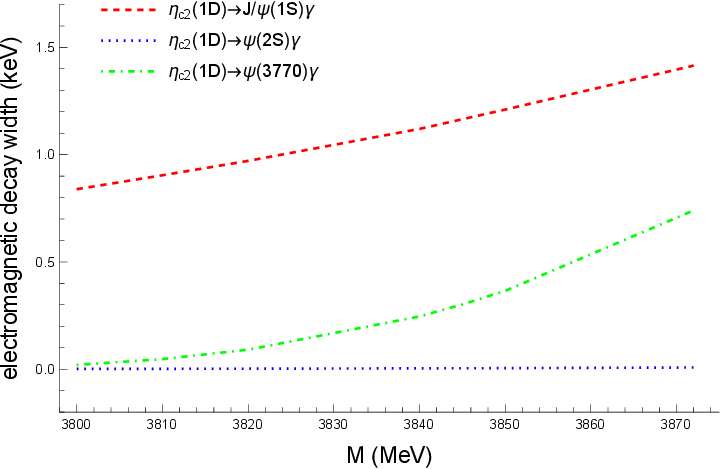}
\end{minipage}
\caption{The total decay width and partial widths of $\eta_{_{c2}}$ vary with the mass $M_{\eta_{_{c2}}}$ .}
\label{mass1}
\end{figure}

{From Figure \ref{mass1}, it can be seen that the decay widths of $\eta_{_{c2}}(1D)\to h_{_{c}}(1P)\gamma$ and $\eta_{_{c2}}(1D)\to\eta_c \pi\pi$ as well as the full width $\Gamma(\eta_{_{c2}})$ are indeed very sensitive to mass $M_{\eta_{_{c2}}}$. The partial decay widths given in Ref. \cite{Jia:2010jn} are the largest in Table \ref{I}, one possible reason is that they took the maximum mass $M_{\eta_{_{c2}}}$ at 3872 MeV.} Taking the dominant channel $\eta_{_{c2}}(1D)\to h_{_{c}}(1P)\gamma$ as an example, when the mass changes from $3800$ MeV to $3872$ MeV, the corresponding width changes from $220$ keV to $407$ keV, and their ratio is
$$
\frac{\Gamma(\eta_{_{c2}}(3800)\rightarrow h_{_{c}}(1P)\gamma)}{\Gamma(\eta_{_{c2}}(3872)\rightarrow h_{_{c}}(1P)\gamma)}=0.541.
$$
Because there are no node structures in the wave functions of initial and final states, this result is purely caused by phase space. Eq.(\ref{Mhc}) shows the transition amplitude of $\eta_{_{c2}}(1D)\rightarrow h_{_{c}}(1P)\gamma$, where the $2\epsilon^{\xi\epsilon_{_{f}}}s_{_{7}}$ term gives the largest contribution. When calculating the square of the amplitude, its power of momentum ${|\vec{P}_{_{f}}}|$ is 4, so we roughly have
$$
\frac{\Gamma(\eta_{_{c2}}(3800)\rightarrow h_{_{c}}(1P)\gamma)}{\Gamma(\eta_{_{c2}}(3872)\rightarrow h_{_{c}}(1P)\gamma)}\propto\frac{{|\vec{P}_{_{f}}}_{3800}|^5}{{|\vec{P}_{_{f}}}_{3872}|^5}=0.459,
$$
this estimate is very close to our calculated value of 0.541. { For the strong decays, $\eta_{_{c2}}(1D)\to\eta_c \pi\pi$ is studied in the framework of the QCD multipole expansion, and its partial width is related to the powers of $M_{\eta_{_{c2}}}-M_{\eta_c}$, so it is also very sensitive to the value of $M_{\eta_{_{c2}}}$; while decay $\eta_{_{c2}}(1D)\to gg$ is only related to mass $M_{\eta_{_{c2}}}$, so within the mass range we choose, its partial width remains almost unchanged.}

\section{CONCLUSION}

In this paper, by constructing the general relativistic wave function for the $2^-$ state according to its quantum number of $J^P$ and solving the complete Salpeter equation of the $2^-$ state, {we obtain the masses  $M_{\eta_{_{c2}}(1D)}=3828.2$ MeV, $M_{\eta_{_{c2}}(2D)}=4157.7$ MeV and $M_{\eta_{_{c2}}(3D)}=4410.7$ MeV, and the decay widths: $\Gamma[\eta_{_{c2}}\rightarrow h_{_{c}}\gamma]=284$ keV, $\Gamma[\eta_{_{c2}}\rightarrow J/\psi\gamma]=1.04$ keV, $\Gamma[\eta_{_{c2}}\rightarrow\psi(2S)\gamma]=3.08$ eV, and $\Gamma[\eta_{_{c2}}\rightarrow\psi(3770)\gamma]=0.143$ keV. With $\Gamma(\eta_{_{c2}}(1D)\to\eta_c \pi\pi)=144~\rm{keV}$ and
$\Gamma(\eta_{_{c2}}(1D)\to gg)= 46.1~\rm{keV}$, the full width of $\eta_{_{c2}}(1D)$ is also estimated and a narrow width of $475$ keV is obtained. We find that the full width strongly depends on the mass of $\eta_{_{c2}}$, if we change the mass from 3800 to 3872 MeV, the total width will change from 369 to 687 keV.}

We also study the contributions of different partial waves which are the nonrelativistic and relativistic terms in the EM decays of $\eta_{_{c2}}$. For the decay $\eta_{_{c2}}\rightarrow h_{_{c}}\gamma$, the main contribution comes from the nonrelativistic dominant partial waves, providing the $E1$ decay, and our calculation includes the contribution of transitions $E1+M2+E3$.
For transition $\eta_{_{c2}}\rightarrow \psi\gamma$, our calculation includes $M1+E2+M3+E4$ transitions, and the relativistic $E2$ transition usually provides the main contribution.

{\bf Acknowledgments}
This work was supported in part by the National Natural Science Foundation of China (NSFC) under the Grants Nos. 12075073, 12375085, 12005169, and 12105072. Q. Li is also supported by the Natural Science Basic Research Program of Shaanxi (Program no. 2021JQ-074), and the Fundamental Research Funds for the Central Universities. B. Wang is also supported by the Start-up Funds for Young Talents of Hebei University (No. 521100221021).


 \end{document}